# COUPLING METHODOLOGY WITHIN THE SOFTWARE PLATFORM ALLIANCES

**Philippe Montarnal** [(1)], Alain Dimier [(2)], Estelle Deville [(1)], Erwan Adam [(1)], Jérôme Gaombalet [(2)], Alain Bengaouer [(1)], Laurent Loth [(2)], Clément Chavant [(3)]

(1) CEA Saclay, 91191 Gif sur Yvette, France
e-mail: philippe.montarnal@cea.fr, estelle.deville@cea.fr, erwan.adam@cea.fr, alain.bengaouer@cea.fr

(2) ANDRA, 1-7 rue Jean Monnet, 92298 Châtenay-Malabry Cedex, France
e-mail : alain.dimier@andra.fr, jerome.gaombalet@cea.fr, laurent.loth@andra.fr

(3) EDF, 1 av. du G. de Gaulle, 92141 Clamart, France
e-mail : clement.chavant@edf.fr

**Abstract:** *CEA, ANDRA and EDF are jointly developing the software platform ALLIANCES which aim is to produce a tool for the simulation of nuclear waste storage and disposal repository. This type of simulations deals with highly coupled thermo-hydro-mechanical and chemical (T-H-M-C) processes. A key objective of Alliances is to give the capability for coupling algorithms development between existing codes. The aim of this paper is to present coupling methodology use in the context of this software platform.*

## 1   INTRODUCTION

The safety assessment of nuclear waste disposals needs to predict coupled thermo-hydro-mechanical and chemical (T-H-M-C) processes, involving phenomena such as heat generation and transport (due to radioactive decay of nuclear waste), infiltration of groundwater (hydrological processes), swelling pressure of buffer material due to saturation (mechanical processes) and chemical evolution of buffer material and porewater (chemical processes). It appeared as necessary to develop and assess numerical tools that model these physical mechanisms.

Therefore, the French Atomic Energy Commission (CEA), and the French Agency for Radioactive Waste Management (ANDRA) have been jointly developing since 2001 the software platform ALLIANCES [1]. The French Electric company (EDF) joined the project in 2003. The aim of the project is to obtain a numerical platform enabling the simulation of all phenomena governing storage and disposal safety as:
-   Efficient coupling of different numerical codes;
-   Simulation of multi-physical and multi-scale phenomena;
-   Uncertainties analysis related to data and models;
-   Studies management and traceability.

Ph. Montarnal, A. Dimier, E. Deville, E. Adam, J. Gaombalet, A. Bengaouer, L. Loth, C. Chavant

More precisely, the necessary physical models are:
- Flow in unsaturated and saturated porous media;
- Radionuclide transport in saturated media (involving convection/diffusion/dispersion transport, sorption via retardation coefficient and radioactive decay, precipitation/dissolution);
- Chemistry/Transport coupling in saturated media with feedback of porosity changes on transport and flow properties ;
- Thermo- Hydraulic and Mechanical coupled processes (THM);
- Thermo-Aerolics;
- Alteration of waste packages and their interaction with its environment.

These models lead to a large set of non-linear differential equations that can be solved on 3D meshing by using finite element like methods

Alliances goal is not to develop a new scientific tool but to accumulate within the same simulation environment the already acquired knowledge and to gradually integrate new ones. Therefore, it is based on the following three fundamental choices:
- Seamless integration of legacy codes as software components ;
- Allow efficient coupling between components by sharing a common data model for mesh and field;
- Sharing open source components for pre and post processing.

The objectives of our coupling strategy are the following:
- Give to physical modellers access to these coupling algorithms ;
- Improve results quality evaluation by using various association of codes ;
- Obtain the best coupling accuracy and cpu cost keeping a code coupling approach.

In the Alliances context, the Chemistry/Transport coupling algorithm is currently used, enabling cross-coupling with different codes [3]: Phreeqc and Chess for the chemical part and Cast3m, MT3D and Traces for the transport part. Other coupled applications will be made available in 2005, canister degradation and corrosion coupled with multi dimensional reactive transport.

We will first describe the software architecture of Alliances. Thereafter, the coupling strategy will be outlined followed by an example of chemistry/transport coupling. Finally, teachings will be drawn and perspectives within the Alliances project will be presented.

Ph. Montarnal, A. Dimier, E. Deville, E. Adam, J. Gaombalet, A. Bengaouer, L. Loth, C. Chavant

## 2 A MULTI-LEVEL ARCHITECTURE

To fulfil the objectives assigned to Alliances, a multi-level architecture has been chosen (see figure 1)

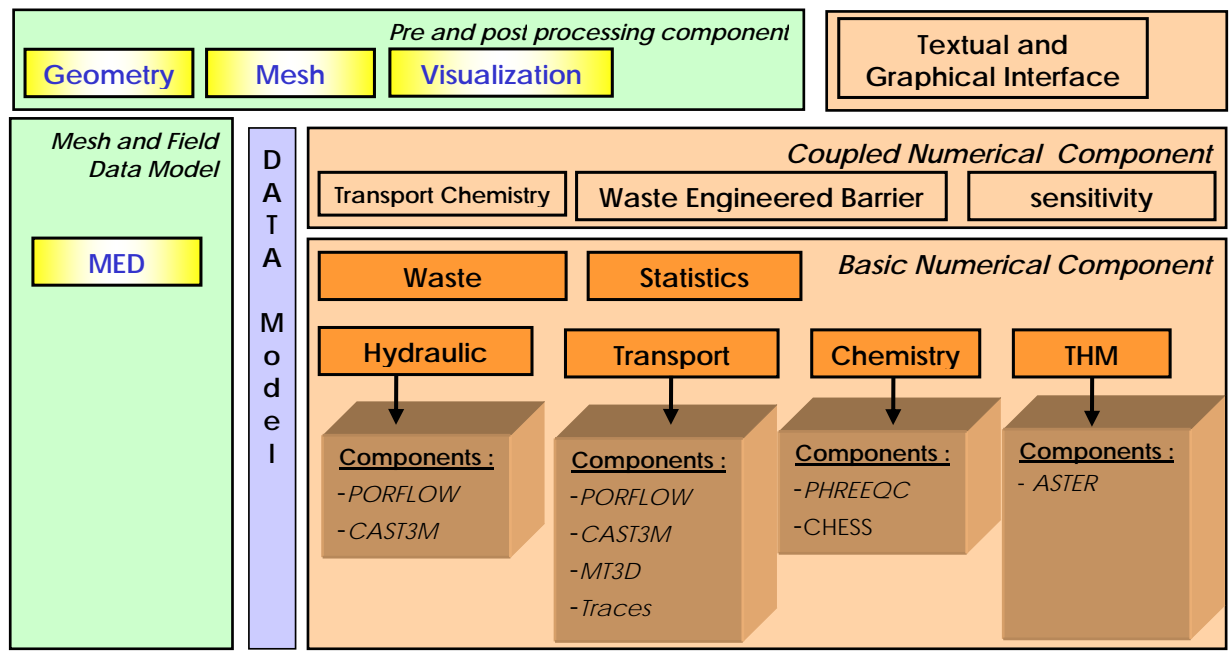

**Figure 1 : Alliances multi-level architecture**

The **basic numerical components** come from the integration of legacy external codes. Each component corresponds to a specific application (Hydraulic, transport, thermo-hydro-mechanic, etc. ). This application can result from coupled processes but the coupling is in this case internal to the integrated code. A component satisfies a given programming interface but may have several implementations (i.e. link to different codes), this property is widely used, for example a transport calculation can be performed with Cast3m or Traces, the choice between one or the other implementation is only specified by a limited set of parameters. These components can be used in a stand-alone way or via coupled numerical components (see next section).

**The coupled numerical components** correspond to multi-physics applications which are implemented via coupling algorithms between the basic numerical components (see part 3)

The **common data model** supports the communication way between the different components. This data model contains all the characteristic values (geometrical, physical, and numerical) used for the modelling and the simulation. The data model joins the physical properties to the geometrical characteristics. This data model is based upon the MED format for the mesh and field part (see next section).

A specific **mesh and field data model** is used. It is based on **MED structure,** a format developed by CEA and EDF in the context of the Salome project [2]. Both file and memory formats are available. File format uses HDF5 (Hierarchical Data Format) technologies allowing efficient, compact and portable data storage. Different tools for basic fields manipulation have been developed (addition, norm computation, ...). Each integrated code can exchange with this format via specific drivers.

The use of Salome platform [2] gives access to the **pre and post processing tools**: geometry and meshing generator, visualization.

The **user interface** is based on a textual mode using Python scripting and a graphical mode using QT and VTK instructions.

**3 COUPLING STRATEGY**

In Alliances, for some applications, as THM, the integrated codes already implement the coupling algorithms. But for other applications, as chemistry/transport or waste package / environment, the coupling should be done at the platform level.

For this reason a key objective of Alliances is to give the capability for coupling algorithms development between existing codes.

The choice of Python was done as central language rather than PVM or MPI.

First of all, it provides an access via an high level language to the components programming interface, by allowing scientific users an easy prototyping of coupling algorithm even if they are not expert in low level programming such as C++.

Moreover, in order to optimize cpu cost, the data exchange needs to be made through memory. For this reason codes are integrated in the platform as dynamic libraries, Python being well adapted to this type of wrapping. More precisely each code appears in Alliances as a Python classes with specific methods for initialization, computation, data exchange. The codes are themselves written in different languages (C, C++, Fortran , Fortran 90). There are several tools to help in Python wrapping. In Alliances we mainly use swig for C and C++ codes and F2PY for Fortran ones.

As previously mentioned several codes can be used for the same physical application. For example, in the chemistry/transport coupling two codes are used for the chemistry part due to model complementarities in the codes.

The algorithms are written at a global level independently from involved codes specificities in order to facilitate the evolution of platform (integration of new codes, coupling algorithms optimisation). Each code share the same Application Programming Interface (API) (see figure 2).

Ph. Montarnal, A. Dimier, E. Deville, E. Adam, J. Gaombalet, A. Bengaouer, L. Loth, C. Chavant

Thanks to the efficient coupling implementation, different algorithms leading to explicit or implicit coupling can be performed. The only limitation being the ability of a component to provide necessary information (derivatives for example). The usual way is to use for weak coupling sequential iterative strategy mainly based upon an operator splitting approach. For stronger coupling, specific work needs to be done in order to develop adapted algorithms [4].

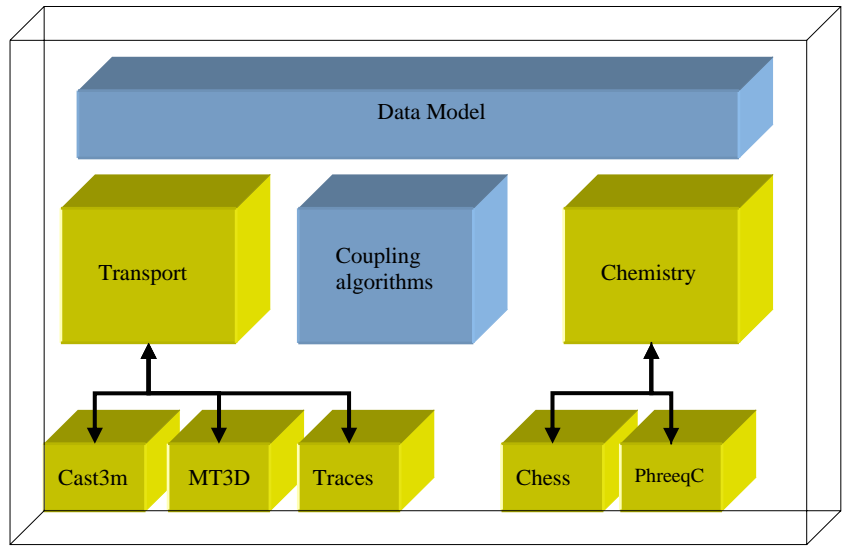

**Figure 2 : coupling stategy, example of chemistry/transport**

## 4 ILLUSTRATION ON CHEMISTRY-TRANSPORT COUPLING

One main application in Alliances of this methodology concerns chemistry/transport coupling[3]. The model that we have to deal with describes the spatial and temporal evolution of a set of chemical species that are, on one hand, submitted to transport phenomena and, on the other hand, submitted to chemical reactions.

We use for the geochemical part CHESS, developed by CIG (France) [5] and PHREEQC, developed by USGS (US) [6] and for hydraulic flow and transport: Cast3M, developed by CEA (France) [7,8] , Modflow/MT3D, developed by U. Alabama (US) [9] and Traces, developed by IMFS (France).

The implementation already done in Alliances uses a sequential iterative algorithm [10,3]. An important validation process was done in order to validate the software against numerous configurations. It is now successfully used to model and simulate real configurations of reactive transport [11] (see figure 3)

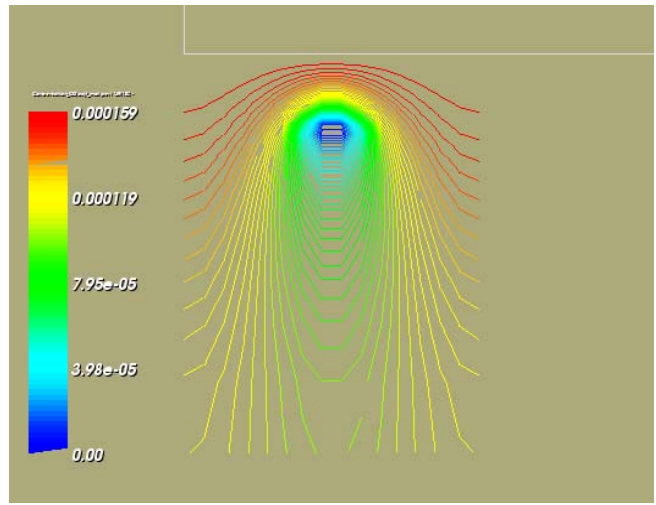

**Figure 3: Oxidative dissolution of uranium dioxide, O2 after 150 years**

## 5 CONCLUSIONS AND PROSPECTS

The coupling approach that we propose here has allowed us to build up rapidly an efficient simulation platform with multi-physics applications. The use of Python and Swig gives access to seamless integration of legacy codes making full benefits of modular developments and maintenance processes. In order to enable an efficient integration, and thereafter use various codes for the same application, we gave a specific attention to the development of algorithms, making them as independent as possible to codes specificities

In Alliances, the Chemistry/Transport coupling algorithm is currently used, enabling cross-coupling with different codes. An other coupled application will be available in 2005: 0D models (canister degradation and corrosion) coupled with multi dimensional reactive transport.

We currently work on coupling algorithms optimization and parallel implementations.

Ph. Montarnal, A. Dimier, E. Deville, E. Adam, J. Gaombalet, A. Bengaouer, L. Loth, C. Chavant